\documentclass[letter]{aa} 
\usepackage{graphicx}
\usepackage{txfonts}
\usepackage{natbib}
\usepackage{wasysym}
\begin{document}
   \title{Non-standard s process in low metallicity massive rotating stars}


   \author{U. Frischknecht\inst{1,2},
           R. Hirschi\inst{2,3},
           F.-K. Thielemann\inst{1}
          }

   \authorrunning{Frischknecht et al.}

   \institute{Department of Physics, University of Basel, Klingelbergstrasse 82, 4056 Basel, Switzerland\\
              \email{urs.frischknecht@unibas.ch}
             \and
               Astrophysics Group, EPSAM, University of Keele, Lennard-Jones Labs, Keele ST5 5BG, UK
             \and
              Institute for the Physics and Mathematics of the Universe, University of Tokyo, 5-1-5 Kashiwanoha, Kashiwa 277-8583, Japan
             }

   \date{}

  \abstract
   {Rotation is known to have a strong impact on the nucleosynthesis of light elements in massive stars, mainly by inducing mixing in radiative zones. In particular, rotation boosts the primary nitrogen production, and models of rotating stars are able to reproduce the nitrogen observed in low-metallicity halo stars.    
   }
   {Here we present the first grid of stellar models for rotating massive stars at low metallicity, where a full s-process network is used to study the impact of rotation-induced mixing on the neutron capture nucleosynthesis of heavy elements. 
   }
   {We used the Geneva stellar evolution code that includes an enlarged reaction network with nuclear species up to bismuth to calculate 25~$M_\odot$ models at three different metallicities ($Z = 10^{-3}, 10^{-5}$, and $10^{-7}$) and with different initial rotation rates.
   }
   {First, we confirm that rotation-induced mixing (shear) between the convective H-shell and He-core leads to a large production of primary $^{22}$Ne (0.1 to 1\% in mass fraction), which is the main neutron source for the s process in massive stars. Therefore rotation boosts the s process in massive stars at all metallicities. Second, the neutron-to-seed ratio increases with decreasing $Z$ in models including rotation, which leads to the complete consumption of all iron seeds at metallicities below $Z = 10^{-3}$ by the end of core He-burning. Thus at low $Z$, the iron seeds are the main limitation for this boosted s process. Third, as the metallicity decreases, the production of elements up to the Ba peak increases at the expense of the elements of the Sr peak. We studied the impact of the initial rotation rate and of the highly uncertain $^{17}$O$(\alpha,\gamma)$ rate (which strongly affects the strength of $^{16}$O as a neutron poison) on our results. This study shows that rotating models can produce significant amounts of elements up to Ba over a wide range of $Z$, which has important consequences for our understanding of the formation of these elements in low-metallicity environments like the halo of our galaxy and globular clusters. Fourth, compared to the He-core, the primary $^{22}$Ne production induced by rotation in the He-shell is even higher (greater than 1\% in mass fraction at all metallicities), which could open the door for an explosive neutron capture nucleosynthesis in the He-shell, with a primary neutron source.
   }
   {}


   \maketitle

%

\section{Introduction}
Massive stars in the early universe are different from the ones observed today. Their low metal content makes them more compact and rotate faster than their equivalents found in the Milky Way. This view is supported by observations of an increasing Be/B-type star ratio with decreasing metallicity \citep{2007A&A...472..577M} and by more rapidly rotating, massive stars in the SMC compared to the Milky Way \citep{2008A&A...479..541H}.
Rapidly rotating stellar models at low $Z$ have been calculated by \citet{2006A&A...447..623M} and \citet{2007A&A...461..571H}. In these models, primary nitrogen yields are much larger than in non-rotating models. When yields from these rotating models are used as input in chemical evolution models, a nice fit of the N/O in very metal poor halo stars \citep[see e.g.][]{2005A&A...430..655S} can be obtained \citep{2006A&A...449L..27C}. This provides a strong support for the occurrence of rotation-induced mixing at low $Z$ and for our models. The primary nitrogen production in rotating low-$Z$ stellar models is accompanied by the primary production of other isotopes like $^{13}$C, and especially $^{22}$Ne, which is the neutron source for s process in massive stars \citep[e.g.][and references therein]{2011RvMP...83..157K}. 

A first attempt to assess the impact of rotation on the s-process nucleosynthesis in low-$Z$ massive rotating stars was made by \citet{2008ApJ...687L..95P}, who investigated the impact of primary $^{22}$Ne in a parametrised way. The recent observation of large s-process enhancements in one of the oldest globular clusters in the bulge of our galaxy supports the view that massive stars could indeed also be important sources for these elements \citep{2011Natur.472..454C}, highlighting the need for comprehensive calculations of the s process in low-$Z$ massive rotating stars. In this letter, we present the first such low-$Z$, massive rotating star models including a full s-process network. 

We describe our models in Sect.~\ref{sec:mixing}. The impact of rotation on the s process in massive 25~$M_\odot$ stars at different metallicities is discussed in Sect.~\ref{sec:sproc}. We give our conclusions in Sect.~\ref{sec:conclusions}.


\section{Stellar evolution models}
\label{sec:mixing}
We calculated our models with the Geneva stellar evolution code (GENEC), which is described in detail in \citep{2008Ap&SS.316...43E}. The main improvement brought to GENEC for this work is the integration of a large nuclear reaction network (613 isotopes up to the end of He-burning and 737 from there on). The smaller network is almost identical to the s-process network used by \citet[][see their Table~1]{2000ApJ...533..998T}. GENEC with the enhanced nucleosynthesis network was first used to study the evolution of the abundance of light elements at the surface of main sequence massive stars \citep{2010A&A...522A..39F}.  Since rotation-induced mixing is prime importance in this work, we briefly review here the input physics used. We used the horizontal diffusion coefficient of \citet{1992A&A...265..115Z} and the shear diffusion coefficient from \citet{1997A&A...317..749T}, which is a conservative choice since this prescription includes a strong reduction of mixing across mean molecular weight gradients. 

We used the reaction library (REACLIB) from \citet{2000ADNDT..75....1R}, but with major updates. The charged particle reaction rates from \citet{1999NuPhA.656....3A} were used except for the following reactions: $^{22}$Ne($\alpha$,n) was taken from \citet{2001PhRvL..87t2501J} and the 3$\alpha$-rate from \citet{2005Natur.433..136F}. If available, the neutron captures were taken from the KADoNiS compilation \citep{2006AIPC..819..123D} and the temperature-dependent beta decays from \citet{1987ADNDT..36..375T}. Several reaction fits were downloaded from the JINA-REACLIB website (groups.nscl.msu.edu/jina/reaclib/db).

For $^{17}$O$(\alpha,\gamma)$ and $^{17}$O$(\alpha,n)$ reaction rates, we used the rates of \citet{1988ADNDT..40..283C} (hereafter CF88) and \citet{1999NuPhA.656....3A}, respectively. Their ratio determines the strength of $^{16}$O as a neutron poison and is very uncertain at the moment. Indeed, \citet{1993PhRvC..48.2746D} predicts that the $^{17}$O$(\alpha,\gamma)$ should be a factor of 1000 smaller than the CF88 rate. This huge uncertainty strongly affects the s~process in massive stars at low Z\footnote{Several other rate uncertainties e.g. $^{22}$Ne$(\alpha,\gamma)$ and  $^{22}$Ne$(\alpha,n)$ also affect the results and will be studied in a forthcoming paper.} \citep{2008IAUS..255..297H}, where $^{16}$O is known to be a strong neutron absorber/poison \citep{2000A&A...354..740R}. Recently, two independent groups have measured the $^{17}$O$(\alpha,\gamma)$ rate \citep{2011nuco.confT,2011PhRvC..83e2802B}, but it is not yet clear whether the new rate will be lower than the CF88-rate at the relevant energies (priv. comm. A. Laird). To assess the impact of a decreased rate, we also calculated 25~$M_\odot$ models with the CF88 rate divided by a factor 10, which will probably still be inside the uncertainties of the new measurements in the relevant energy range.

We calculated models of 25~$M_\odot$ stars at initial metallicities, $Z=10^{-3}$, $10^{-5}$, and $10^{-7}$ (the corresponding [Fe/H]-values are given in Table~\ref{tab:models}), to investigate the influence of rotation on the s process in massive stars. At all three metallicities, we assumed an $\alpha$-enhanced composition\footnote{ with the $\alpha$-elements enhanced with respect to iron, i.e. $\mathrm{[X/Fe]}=A=\mathrm{constant}$ (since $\mathrm{[Fe/H]}\le-1$) where $A=+0.562$, $+0.886$, $+0.500$, $+0.411$,$+0.307$, $+0.435$, $+0.300$, $+0.222$, and $+0.251$ for $^{12}$C, $^{16}$O, $^{20}$Ne, $^{24}$Mg, $^{28}$Si, $^{32}$S, $^{36}$Ar, $^{40}$Ca, and $^{48}$Ti, respectively}. All other elements were scaled from the solar composition published in \citet{2005ASPC..336...25A}. At each $Z$, a model without rotation was calculated for reference. For rotating models, we chose $\upsilon_{\rm ini}/\upsilon_{\rm crit}=0.4$ as a standard rotation rate, where $\upsilon_{\rm ini}$ is the initial surface velocity and $\upsilon_{\rm crit}$ the critical rotation velocity, for which the centrifugal force balances the gravity at the equator. This ratio corresponds to an average equatorial velocity of $\langle\upsilon_{\rm eq}\rangle_{\rm MS}\approx214$~km~s$^{-1}$ on the main sequence at $Z_\odot$ in our models. This is slightly lower than the peak velocity of the velocity distribution of B-type stars in our Galaxy found by \citet{2006A&A...457..265D}. A constant $\upsilon_{\rm ini}/\upsilon_{\rm crit}$ leads to a slight decrease in initial angular momentum with decreasing Z and in our models at lowest Z to $\langle\upsilon_{\rm eq}\rangle_{\rm MS}\approx383$~km~s$^{-1}$. At low $Z$, massive stars could have a higher initial angular momentum according to the latest star formation simulations \citep{2011MNRAS.tmp..142S}, so we also calculated models with $\upsilon_{\rm ini}/\upsilon_{\rm crit}=0.5$ for $Z= 10^{-5}$ and $\upsilon_{\rm ini}/\upsilon_{\rm crit}=0.6$ for  $Z=10^{-7}$, corresponding to $\langle\upsilon_{\rm eq}\rangle_{\rm MS}=428$ and $588$~km~s$^{-1}$, respectively. All models were computed until at least the start of O-burning. Beyond this point, the carbon and helium-burning shells do not change significantly, so the s-process yields can be estimated well at this point. The yields can still be slightly modified by late interactions between shells and increased temperature just before the collapse and by explosive nucleosynthesis \citep[see e.g.][]{2009ApJ...702.1068T}.

We list the properties of the models in Table~\ref{tab:models} in the following order: model name (row 1), initial metallicity (2), [Fe/H] (3), $\upsilon_{\rm ini}/\upsilon_{\rm crit}$ (4), average rotation rate on the main sequence $\langle\upsilon\rangle_{\rm MS}$ (5), the mass fraction of $^{22}$Ne burned during core He-burning $\Delta$X$(^{22}$Ne$)_{\rm c}$ (6), the mass fraction of $^{22}$Ne at the start of C-burning, X$(^{22}$Ne$)$ (7), the average number of neutron captures per iron seed $n_c$ (8), and the pre-supernova yields in solar masses and production factors, f, for the elements Sr (9/10), and Ba (11/12). The remnant masses used to calculate the yields \citep[see method in][]{2007A&A...461..571H} range between 2.3 and 2.8 $M_\odot$ and are usually situated close to the bottom of the last carbon shell.

\begin{table*}[hbt]
\caption{ Parameters of the 25 $M_\odot$ models calculated. }
 \begin{tabular}{lccccccccccccc}
  \hline\hline\\[-8pt]
model$^a$					&  A0      &  A1      & B0	 &  B1      &  B2$^b$  &   B3     &   B4$^b$ &    C1    &  C3      &   C4$^b$  \\
$Z$						& $10^{-3}$& $10^{-3}$& $10^{-5}$& $10^{-5}$& $10^{-5}$& $10^{-5}$& $10^{-5}$& $10^{-7}$& $10^{-7}$& $10^{-7}$ \\
$\mathrm{[Fe/H]}$				&  $-1.8$  &  $-1.8$  & $-3.8$	 &  $-3.8$  &  $-3.8$  &  $-3.8$  &  $-3.8$  &   $-5.8$ &  $-5.8$  &  $-5.8$   \\
$\upsilon_{\rm ini}/\upsilon_{\rm crit}$  	&  0       &  0.4     & 0	 &  0.4     &  0.4     &  0.5     &  0.5     &   0.4    &   0.6    &  0.6      \\
$\langle\upsilon\rangle_{\rm MS}^c$ 		&  0       &  285     & 0	 &  333     &  333     &  428     &  428     &  383     &   588    &   588     \\
$\Delta$X$(^{22}$Ne$)^d$  		        &$7.52(-4)$&$4.08(-3)$&$7.21(-6)$&$1.23(-3)$&$1.27(-3)$&$3.83(-3)$&$3.75(-3)$&$1.05(-4)$&$4.57(-3)$&$4.44(-3)$ \\  
X$(^{22}$Ne$)$					&$4.16(-4)$&$6.22(-4)$&$4.14(-6)$&$1.69(-4)$&$1.82(-4)$&$4.94(-4)$&$4.85(-4)$&$3.81(-5)$&$2.68(-4)$&$3.11(-4)$ \\
$n_c^e$						&  1.64    &  12.7    & 0.08	 & 5.77     &   23.1   & 16.5     & 31.8     &  14.0    & 33.5     &  48.5     \\  
  \hline\\[-8pt]
&\multicolumn{9}{c}{pre-SN yields [$M_\odot$]}\\
  \hline \\[-8pt]
Sr 						&$1.66(-8)$&$1.13(-5)$&$2.60(-11)$&$8.70(-9)$ &$3.82(-7)$&$2.05(-7)$ &$4.36(-7)$&$1.08(-9)$ &$3.66(-9)$ &$3.02(-9)$  \\
f(Sr)						& 1.75     & 501.4    & 1.12      &  28.86    & 1724     & 944.7     & 1968     & 489.1     & 1650      & 1361\\
Ba 						&$2.13(-9)$&$2.29(-8)$&$3.56(-12)$&$7.41(-11)$&$6.49(-9)$&$4.89(-10)$&$4.73(-8)$&$3.80(-10)$&$9.86(-10)$&$2.95(-9)$  \\
f(Ba)						& 1.34     &  3.87    & 1.06      &   2.10    & 106.1    &  9.07     & 767.0    & 615.6     & 1598      & 4771\\
  \hline \\
  \end{tabular}
\\
Notes.  
$^{(a)}$ Model names (1$^{\rm st}$ line) include a letter and a number. The letter refers to the initial metallicity, $Z$ (2$^{\rm nd}$ line) and the number to the initial rotation/${17}$O$(\alpha,\gamma)$ rate used. 0 is for non-rotating models, 1 for models with constant $\upsilon_{\rm ini}/\upsilon_{\rm crit}=0.4$, 3 for increasing $\upsilon_{\rm ini}/\upsilon_{\rm crit}: 0.5$-$0.6$ with decreasing $Z$ and 2 and 4 for lower ${17}$O$(\alpha,\gamma)$ rates (CF88/10).
$^{(b)}$ Model with the ${17}$O$(\alpha,\gamma)$ reaction rate of CF88 divided by 10. 
$^{(c)}$ in km~s$^{-1}$.
$^{(d)}$ Since $^{22}$Ne is produced and destroyed at the same time in rotating stars, we derived this value from the sum of the $^{25}$Mg and $^{26}$Mg produced during central He-burning.
$^{(e)}$ Number of neutron captures per initial seed. 
\label{tab:models}
\end{table*}

We can see from Table~\ref{tab:models} ($\Delta$X$(^{22}$Ne$)$) that rotating models at all metallicities produce and burn significant amounts of $^{22}$Ne, confirming the results of previous studies \citep{2006A&A...447..623M,2007A&A...461..571H,2008ApJ...687L..95P}. The amount of primary $^{22}$Ne in the convective He-core at the end of He-burning when s process is active is between 0.1 and 1\% in mass fractions. Considering a constant value of $\upsilon_{\rm ini}/\upsilon_{\rm crit}=0.4$ at all metallicities (models A1, B1, C1), the primary $^{22}$Ne in the He-core decreases slightly with decreasing metallicity. There is, however, theoretical and observational support to consider a slight increase in $\upsilon_{\rm ini}/\upsilon_{\rm crit}$ with decreasing metallicity. Considering models A1 ($\upsilon_{\rm ini}/\upsilon_{\rm crit}=0.4$ at $Z = 10^{-3}$), B3 ($\upsilon_{\rm ini}/\upsilon_{\rm crit}=0.5$ at $Z = 10^{-5}$), and C3 ($\upsilon_{\rm ini}/\upsilon_{\rm crit}=0.6$ at $Z = 10^{-7}$), which correspond to an increase in $\upsilon_{\rm ini}/\upsilon_{\rm crit}$ with decreasing metallicity, we see that rotating models produce and burn a constant quantity of $^{22}$Ne, around 0.5\% in mass fraction. We can see from these results that rotating models produce significant amounts of $^{22}$Ne in a primary way over the entire range of metallicity computed, whereas in non-rotating models (A0, B0), the amount of $^{22}$Ne available is secondary.

In the He-shell, the $^{22}$Ne mass fraction in rotating models is, by a factor of 19 (A1), 1100 (B1) and 13400 (C1) higher than the corresponding non-rotating models. While the $^{22}$Ne in the He-shell of non-rotating model is purely secondary, in rotating models it is primary and amounts to 1 and 2\% in mass fraction at the pre-SN stage and independent of metallicity. This is very interesting for explosive nucleosynthesis in the He-shell. This site was investigated for the r-process by \citet[][]{1978ApJ...222L..63T} and \citet[][]{1979A&A....74..175T}, but later on considered only as a possible site for a weak r-process \citep{2002RvMP...74.1015W} or n-process \citep{2002ApJ...576..323R}. Given the large amounts of primary $^{22}$Ne produced in the rotating models at all $Z$, it is worthwhile reconsidering explosive nucleosynthesis in the He-shell.
    
\section{Non-standard s process}
\label{sec:sproc}
The standard s process in massive stars is a secondary process \citep[see e.g.][]{1992ApJ...387..263R}, because the main neutron source ($^{22}$Ne coming from the initial C, N, and O) and the seeds (mainly iron) have a secondary origin, while neutron poisons are a mixture of primary (mainly $^{16}$O for He-burning and $^{20}$Ne, $^{24}$Mg for C-burning) and secondary (mainly $^{25}$Mg for He-burning) elements. At low $Z$, the contribution from the carbon shell becomes smaller (given the mass fraction of $^{22}$Ne left at the start of C-burning, X$(^{22}$Ne$)$, see line 7 in Table~\ref{tab:models}). In non-rotating models (A0, B0), the standard s process is even less efficient than a secondary process.

Rotation significantly changes the structure and pre-SN evolution of massive stars \citep{2004A&A...425..649H} and thus also the s-process production. Rotating stars have similar central properties to more massive non-rotating stars. In particular they have more massive helium-burning cores and higher central temperature. 
The latter means that the He-core contribution increases at the expense of the C-shell contribution. This and the primary origin of neutron poisons during carbon burning lead to a very small C-shell contribution to the pre-SN yields at low $Z$, i.e. lower than 10\% for all the rotating models presented in Table~\ref{tab:models}. 

\begin{figure*}
 \includegraphics[width=0.90\textwidth,height=0.25\textheight]{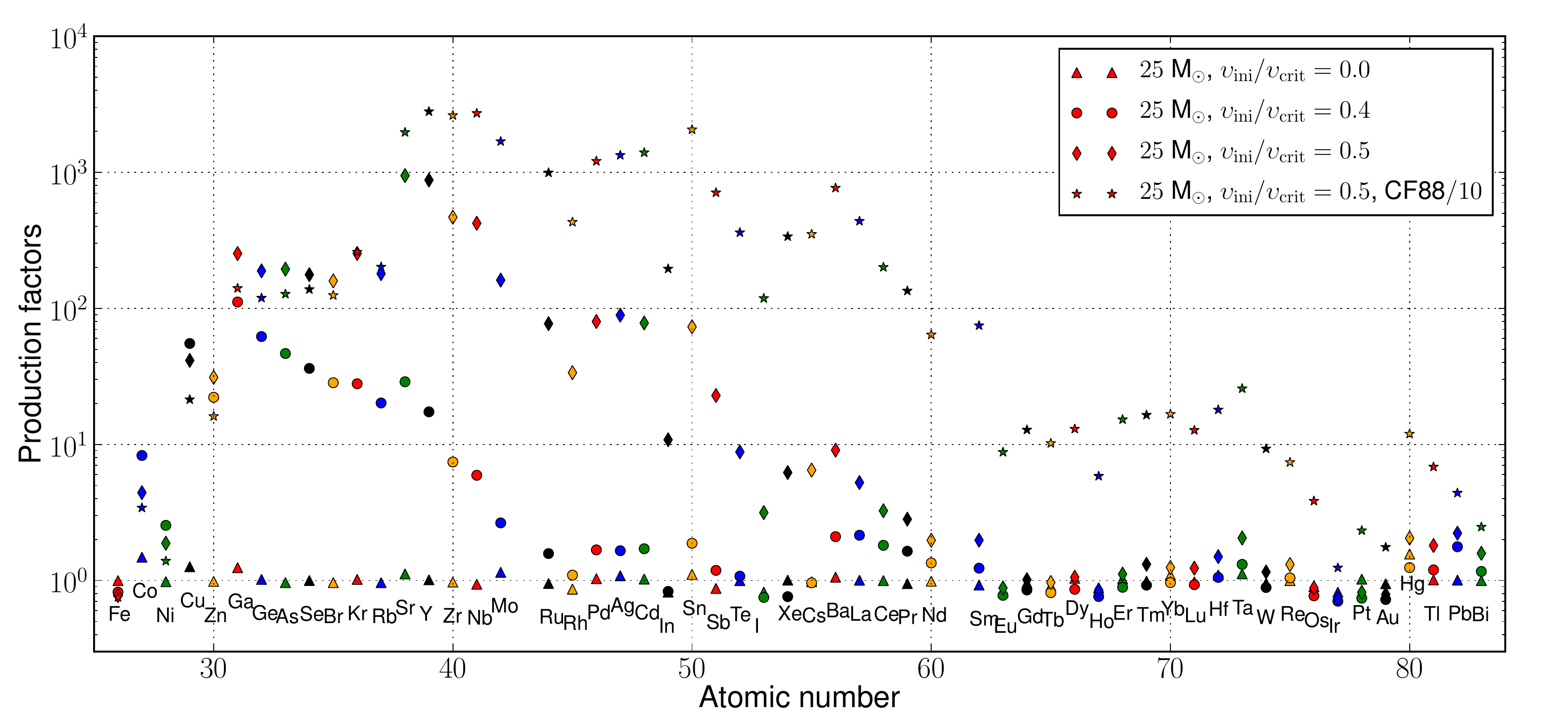}
 \caption{Production factors (ejected mass divided by the initial mass of the element) for the 25~$M_\odot$ models with $Z=10^{-5}$. The model without rotation (triangles) does not produce s-process efficiently, whereas the rotating models (circles, B1, and diamonds, B3) do. The additional, rotating models with reduced $^{17}$O$(\alpha,\gamma)$ rates (stars, B4, CF88/10) highlight the uncertainty for the neutron poison $^{16}$O.
}
 \label{fig:sproc} 
\end{figure*}

All the effects of rotation boost the s-process production at all metallicities compared to non-rotating models as can be seen in Table~\ref{tab:models}. At $\mathrm{[Fe/H]}=-1.8$ (models A0 and A1), the rotating model burns about five times more $^{22}$Ne than the non-rotating model during core He-burning. Although a factor of 5 is not very large, 96\% of the initial iron is transformed into s-process elements in the rotating model, whereas only 53\% of the initial iron is transformed in the non-rotating model. In the rotating model, the neutron source is sufficient to deplete all the seeds, and the production is limited by the seeds (not the neutron source any more). This explains why the s-process yields are much larger in the rotating model, by up to a factor of 1000. Nevertheless, at $\mathrm{[Fe/H]}=-1.8$, rotating models produce elements in large quantities only up to the Sr peak. We thus expect that the non-standard s~process is qualitatively similar to the normal s~process (production of elements with $A=60$-$90$) for $\mathrm{[Fe/H]}>-2$ but with higher production factors. 
At $\mathrm{[Fe/H]}=-3.8$ (see Fig.~\ref{fig:sproc} showing production factors at the pre-SN stage), elements up to Sr are still strongly produced in the rotating model (B1, Fig.~\ref{fig:sproc}), whereas the overproduction is very small in the non-rotating model. Depending on the initial rotation rate and the $^{17}$O$(\alpha,\gamma)$ rate used, elements up to Ba start to be efficiently produced (see discussion below).
At $\mathrm{[Fe/H]}=-5.8$, both the rotating and non-rotating models produce and burn enough $^{22}$Ne to deplete all the iron seeds. At this very low $Z$, the main limitation is the seeds and, given the low initial iron content, the s-process production remains very modest.

It is interesting to look at the metallicity dependence of the non-standard s-process production in rotating models. As said, the production is limited mainly by the iron seeds. Even at the lowest metallicities in a very rapidly rotating model (C3 and C4, $\upsilon_{\rm ini}/\upsilon_{\rm crit}$= 0.6 instead of the standard 0.4), and thus with a larger primary neutron source, there is no additional production of s-process elements starting from light element seeds like $^{22}$Ne. Instead, what happens is that not only iron is depleted, but also elements up to Sr are partially destroyed (after being produced) and heavier elements like Ba are produced ($\mathrm{[Sr/Ba]}\le0$). 
Indeed, going from $\mathrm{[Fe/H]}=-3.8$ (B1) to $\mathrm{[Fe/H]}=-5.8$ (C1), the Sr yield decreases by a factor of $\approx 9$, while the Ba yield increases by a factor of 5. We therefore have a different metallicity dependence for the production of elements belonging to the different peaks: there is a roughly secondary production of elements up to Sr but the $Z$ dependence for heavier elements like Ba is milder. The secondary-like behaviour of Sr/Y/Zr in the metallicity range covered by our models, makes the non-standard s process in massive rotating stars an unlikely solution for the LEPP (light element primary process) problem at low $Z$ \citep{2004ApJ...601..864T}.

Apart from the seeds, the s process strongly depends on the neutron source and neutron poisons. Both are still quite uncertain at present. The neutron source, $^{22}$Ne, depends on rotation-induced mixing, hence on the initial velocity of stars at very low $Z$. To assess the uncertainty linked to the primary production of $^{22}$Ne, we calculated additional models at $\mathrm{[Fe/H]}=-3.8$ and $-5.8$ with a higher initial velocity (B3/B4 $\upsilon_{\rm ini}/\upsilon_{\rm crit}=0.5$ and C3/C4, $\upsilon_{\rm ini}/\upsilon_{\rm crit}=0.6$). Compared to the C1 case ($\upsilon_{\rm ini}/\upsilon_{\rm crit}$= 0.4), the increase in the production of $^{22}$Ne is about a factor of 4. This leads to a higher neutron capture per seed ($n_c$) and thus to a production of elements like Ba at the expense of elements like Ge, but the total production (sum of all isotopes heavier than iron) is still limited by the iron seeds, as said above. A major uncertainty concerning neutron poisons is the importance of $^{16}$O as a neutron poison. As explained in the previous section, at low $Z$, $^{16}$O is a strong neutron absorber during core He-burning. The neutrons captured by $^{16}$O$(n,\gamma)^{17}$O may either be recycled via $^{17}$O$(\alpha,n)^{20}$Ne or lost via $^{17}$O$(\alpha,\gamma)^{21}$Ne. To investigate the impact of the $^{17}$O$(\alpha,\gamma)^{21}$Ne uncertainty, we calculated models with the CF88 rate divided by a factor of 10 (B2, B4, and C4). The impact of a change of even a factor of 10 in this rate is strong (see Table~\ref{tab:models} and Fig.~\ref{fig:sproc}). Given the differences between models with the CF88 and CF88/10 rate, the experimental determination of the $^{17}$O$(\alpha,\gamma)$-rate and the $^{17}$O$(\alpha,n)$ is crucial to a more accurate prediction for the s process in massive rotating stars at low metallicity.

Characteristic quantities for the s process are [Sr/Ba] and [Sr/Pb]. For model A1 at $Z=10^{-3}$, $\mathrm{[Sr/Ba]}=+2.33$, which is typical of the standard s process in massive stars. On the other hand, the models at lower metallicities show reduced values between $+2.05$ and $+0.42$  at $Z=10^{-5}$, and between $+0.02$ and $-0.54$ at $Z=10^{-7}$ (although the $Z=10^{-7}$ models will not contribute much to the galactic chemical enrichment). We therefore generally obtain $\mathrm{[Sr/Ba]}>0$ from the low-$Z$, massive rotating star, s~process. Pb is only produced in very extreme cases and with $\mathrm{[Pb/Sr]}\apprle-1$. Thus massive stars cannot be responsible for CEMP-s stars. The models of \citet{2008ApJ...687L..95P} efficiently produce elements up to Pb. This difference between the two studies is explained by the fact that \citet{2008ApJ...687L..95P} used the $^{17}$O$(\alpha,\gamma)$ of \citet{1993PhRvC..48.2746D}, which is now disfavoured by recent experiments. Finally, a similar spread for [Sr/Ba] is obtained as the one used in \citet{2011Natur.472..454C} to explain the abundance of one of the oldest globular clusters in the galactic bulge, thus supporting the conclusions of that study.

\section{Conclusions}
\label{sec:conclusions}
We have calculated the first complete stellar models including a full s-process network to study the effects of rotation on the s process in massive stars in low-metallicity environments. Our models confirm that rotation-induced mixing in radiative zones leads to a primary production of $^{22}$Ne in the He core over the wide $Z$ range studied ($\mathrm{[Fe/H]}=-5.8$, $-3.8$, and $-1.8$). From $0.1$ to $1$\% of $^{22}$Ne in mass fraction is produced in central He-burning. The large primary production of $^{22}$Ne also occurs in the He-shell in all the rotating models calculated, i.e. 1-2\% in mass fraction at all $Z$. This can have a strong impact on the supernova nucleosynthesis in He-burning shell, since it could provide a primary neutron source for neutron capture nucleosynthesis at all values of Z. We investigated the impact of the $^{17}$O$(\alpha,\gamma)$-rate on the strength of $^{16}$O as a neutron poison and find the impact of current uncertainties still significant, thus deserving further experimental studies.

The s process in massive stars with rotation contributes significantly to the production of elements up to Sr above $\mathrm{[Fe/H]}\approx-2$, the main limiting factor being the iron seeds at low $Z$. The production of elements up to Ba is sensitive to the initial rotation rate, and we generally obtain $\mathrm{[Sr/Ba]}>0$ from low-$Z$ massive rotating star s process. Pb is only produced in very extreme cases, but with $\mathrm{[Pb/Sr]}\apprle-1$ massive stars cannot be responsible for CEMP-s stars.
A similar spread for [Sr/Ba] to the one used in \citet{2011Natur.472..454C} was obtained to explain the unique abundance of several stars in one of the oldest globular clusters in the galactic bulge, thus supporting the conclusions of that study. Further stellar and galactic chemical evolution models will assess the full impact of this boosted s~process.

\begin{acknowledgements}
The authors gratefully acknowledge fruitful discussions with and comments from Marco Pignatari and Georges Meynet. This work was supported by the SNSF and by the ESF EUROCORES programme EuroGENESIS. RH and UF acknowledge support from the Royal Society (IJP090091 grant).
RH acknowledges support from the WPI Research Center Initiative, MEXT, Japan. 
\end{acknowledgements}


\end{document}